\documentclass{amsart}
\usepackage{amssymb}
\usepackage[footnotesize,bf]{caption}

\theoremstyle{definition}

\theoremstyle{remark}

\def\beq{\begin{eqnarray}}
\def\eeq{\end{eqnarray}}
\def\bsp{\begin{split}}
\def\esp{\end{split}}

\def\d{\mathrm{d}}

\newcommand{\fvec}[4]{{#1}_{\{a}{#2}_b{#3}_c{#4}_{d\}}}

\newcommand{\lnlm}[1]{\fvec{\ell}{n}{\ell}{m^{#1}}}
\newcommand{\lmlm}[2]{\fvec{\ell}{m^{#1}}{\,\ell}{m^{#2}}}

\newcommand{\mmlm}[3]{\fvec{m^{#1}}{m^{#2}}{\,\ell}{m^{#3}}}

\newcommand{\be}{\begin{equation}}
\newcommand{\ee}{\end{equation}}
\newcommand{\msub}[2]{m^{#1}{}_{#2}}

\def \bl {\mbox{\boldmath{$\ell$}}}

\def \bn {\mbox{\boldmath{$n$}}}

\def \hbm #1 {\mbox{\boldmath{$\hat m^{(#1)}$}}}
\def \bm {\mbox{\boldmath{$m$}}}

\newcommand{\bo}{{\mathcal{B}}}

\def \bk {\mbox{\boldmath{$k$}}}

\begin{document}
\hspace{9cm} NIKHEF/2006-008
\vspace{1cm}
\title{\textbf{Higher Dimensional $VSI$ Spacetimes}}
\author{\textbf{A. Coley, A. Fuster, S. Hervik and N. Pelavas}}

\address{Department of Mathematics and Statistics,
Dalhousie University, Halifax, Nova Scotia, Canada B3H 3J5 (AC, SH and NP); National Institute for Nuclear and High-Energy
Physics (NIKHEF), Kruislaan 409, 1098 SJ, Amsterdam, The Netherlands (AF)}
\email{fuster@nikhef.nl; aac,herviks,pelavas@mathstat.dal.ca}

\date{\today}
\maketitle

\begin{abstract}

We present the explicit metric forms for higher dimensional 
vanishing scalar invariant ($VSI$) Lorentzian spacetimes. 
We note that all of the $VSI$
spacetimes belong to the higher
dimensional Kundt class.
We determine all of the $VSI$ spacetimes which admit a covariantly constant
null vector, and we note that in general in higher dimensions these 
spacetimes are of Ricci type $III$ and Weyl type $III$. The Ricci type $N$ subclass is related to the 
chiral null models and includes the relativistic gyratons and the higher dimensional pp-wave spacetimes. 
The spacetimes under investigation are of particular interest since they are solutions of supergravity or
superstring theory.

\end{abstract}
\vspace{1cm}
\noindent 
[PACS: 04.20.Jb, 04.65+l]

\vskip .1in

\vskip .1in

\section{Introduction}

Recently \cite{Higher} it was proven that all curvature invariants
of all orders vanish in an $N$-dimensional Lorentzian ($VSI$) spacetime if
and only if there exists an aligned shear-free, non-expanding, non-twisting,
geodesic null direction $\ell^a$ along which the Riemann tensor
has negative boost order. An analytical form of these conditions
are as follows: \begin{equation} R_{abcd} = 8 A_{i} \, \lnlm{i}+ 8
B_{ijk} \, \mmlm{i}{j}{k}+ 8 C_{ij} \, \lmlm{i}{j} \label{RRR}
\end{equation}  where $i= 1... (N-2)$ (i.e., the Riemann tensor is of algebraic type III
N or O \cite{class}), and
\begin{equation} \ell_{a ; b} = L_{11} \ell_a \ell_b  + L_{1i} \ell_a
\msub{i}{b} + L_{i1} \msub{i}{a} \ell_b; \label{l_ab-VSI}
\end{equation} that is, the expansion matrix and the twist
matrix are zero (as well as $L_{i0}=0=L_{10}$, corresponding to an
affinely parametrized geodesic congruence $\ell_a$).

This result generalizes a previous theorem in four dimensions (4D) \cite{4DVSI}.
Indeed, in 4D $VSI$ spacetimes can be classified according to their
Petrov type, Segre type and the vanishing or non-vanishing of the
spin coefficient $\tau$;  all of the corresponding metrics are
displayed in \cite{4DVSI}.

A number of higher dimensional $VSI$ spacetimes are explicitly known
\cite{CMPPPZ,frolov}. However, we wish to complete this investigation and write down
the metric for higher dimensional $VSI$ spacetimes in a canonical
form (similar to what is done in 4D). We note that all of the VSI
spacetimes have a shear-free, non-expanding, non-twisting geodesic
null congruence $\bl=\partial_v$, and hence belong to the higher
dimensional {\em Kundt} class \cite{CMPPPZ,CSI}. In this paper, we shall therefore present 
the explicit metric forms for higher
dimensional $VSI$ spacetimes \cite{CSI}.

\subsection{Preliminaries}

We shall consider a null frame   $\bl=\bm_0,\ \bn=\bm_1,\ \bm_2,
...\ \bm_{N-1}$ ($\bl,\ \bn$ null with  $\ell^a \ell_a= n^a n_a =
0$, $\ell^a n_a = 1$, $\bm_i$ real and spacelike $m_i{}^a m_j{}_a
= \delta_{ij}$, $i=2,...,N-1$,  all other products vanish) in an
$N$-dimensional Lorentz-signature spacetime, so that
\begin{equation} g_{ab} = 2\ell_{(a}n_{b)} +  \delta_{jk} m^j{}_a
m^k{}_b. \label{tetrad}
\end{equation}

A {\em null rotation} about $\bn$ is a Lorentz transformation of
the form
\begin{equation}
  \label{eq:nullrot}
    \hat{\bn}=  \bn,\quad
    \hat{\bm}_i=  \bm_i + z_i \bn,\quad
    \hat{\bl}= \textstyle \bl
    -z_i \bm^i-\frac{1}{2} \delta^{ij} z_i z_j\, \bn.
\end{equation}
A null rotation about $\bl$ has an analogous form.  A boost is a
transformation of the form
\begin{equation}
  \label{eq:boost}
    \hat{\bn}= \lambda^{-1}\bn,\quad
    \hat{\bm}_i=  \bm_i,\quad
    \hat{\bl}=  \lambda\, \bl, \quad \lambda \neq 0.
\end{equation}
Using the notation 
\begin{equation} w_{\{a} x_b y_c z_{d\}} \equiv \frac{1}{2}(w_{[a} x_{b]}
y_{[c} z_{d]}+
 w_{[c} x_{d]} y_{[a} z_{b]}) \equiv \frac{1}{8}\{[w_p x_q] [y_r z_s]\} ,\label{zavorka}
\end{equation}
we can decompose the Weyl tensor $C_{abcd}$ and sort its components by
boost weight (see Table 1 in \cite{Algclass}). The Weyl
scalars also satisfy a number of additional relations, which follow
from curvature tensor symmetries and from the trace-free
condition.

The boost order of a tensor is a function of the null
direction $\bk$. We denote boost order by $\bo(\bk)$.  We will
call the integer $1-{\bo(\bk)}\in \{0,1,2,3\}$ the {\em order of
alignment}. The  principal type of a Lorentzian manifold
  is I, II, III, N according to whether there exists an aligned $\bk$
  of alignment order $0,1,2,3$, respectively (i.e.,
  ${\bo(\bk)}=1,0,-1,-2$, respectively) \cite{Algclass,class}.  If no aligned $\bk$
  exists we will say that the manifold is of type G.  If the Weyl
  tensor vanishes, we will say that the manifold is of type O.
It follows that there exists a frame in which the components of
the Weyl tensor satisfies:
\begin{eqnarray}
&&  Type\ I:\ \ ~~ C_{0i0j}=0, \nonumber\\
&&  Type\ II: \ ~~C_{0i0j}=C_{0ijk}=0, \nonumber \\
 && Type\ III:  ~~C_{0i0j}=C_{0ijk}=C_{ijkl} =C_{01ij}=0, \nonumber\\
 && Type\ N:  \ ~~C_{0i0j}=C_{0ijk}=C_{ijkl} = C_{01ij}=C_{1ijk}=0.
\end{eqnarray}
The general types have various algebraically special subtypes
\cite{class}; for example, Type III(a) for $C_{011i}=0$.

\section{Higher dimensional $VSI$ metrics}

From \cite{CMPPPZ,CSI}, it follows that
any {$VSI$} metric can be written in the form \beq \d s^2=2\d u\left[\d
v+H(v,u,x^n)\d u+W_{i}(v,u,x^n)\d x^i\right]+\delta_{ij}\d x^i\d
x^j \label{Kundt}\eeq with $i,j=1,\dots,N-2$. The~metric functions $H$ and $W_{i}$
satisfy the remaining vanishing scalar invariant conditions and
the~Einstein equations.
It is convenient to introduce the null frame \beq
\ell&=& \d u, \\
{\bf n}&=& \d v+H\d u+W_{ i}{{\bf m}}^{ i+1}, \\
{\bf m}^{ i+1}&=& \d x^i. \eeq

Spacetimes possessing a null vector field $\ell$ obeying \beq
\ell^{A}\ell_{B;A}=\ell^{A}_{~;A}=\ell^{A;B}\ell_{(A;B)}=\ell^{A;B}\ell_{[A;B]}=0; \eeq i.e.,
$\ell$ is geodesic, non-expanding, shear-free and non-twisting (i.e., $L_{ij} \equiv \ell_{i;j}=0$
\cite{Higher}), are higher-dimensional \emph{Kundt
metrics} (or simply Kundt metrics), since they generalise the 4-dimensional Kundt metrics.
There is some remaining freedom in the choice of frame; 
namely null rotations about $\ell$,
spins and boosts. 

The generalized Kundt metrics are given by (\ref{Kundt}), but with a general
transverse metric of the form $g_{ij}(u,x^k)$. From \cite{CSI}, it follows that
for any {$VSI$} member of the generalized Kundt class, there is a transformation of
the form: 
\beq
(v',u',x'^i)=(v,u,f^i(u;x^k)), \label{trans} \eeq
which transforms the transverse metric  $g_{ij}(u,x^k)$
to flat space  with $\delta_{ij}$. Hence the {$VSI$} spacetimes are Kundt spacetimes
of the form (\ref{Kundt}).

The remaining coordinate
freedom preserving the Kundt form are:
\begin{enumerate}
\item{} $(v',u',x'^i)=(v,u,f^i({x}^k))$ and $J^i_{~j}\equiv \frac{\partial f^i}{\partial x^j}$.
\beq
H'= H, \quad W'_i=W_j\left(J^{-1}\right)^j_{~i}, \quad \delta'_{ij}=\delta_{kl}\left(J^{-1}\right)^k_{~i}\left(J^{-1}\right)^l_{~j}.\nonumber
\eeq
\item{} $(v',u',x'^i)=(v+h(u,x^k),u,x^i)$
\beq
H'=H-h_{,u}, \quad W'_i=W_i-h_{,i}, \quad \delta_{ij}'=\delta_{ij}.\nonumber
\eeq
\item{} $(v',u',x'^i)=(v/g_{,u}(u),g(u),x^i)$
\beq
H'=\frac 1{g_{,u}^2}\left(H+v\frac{g_{,uu}}{g_{,u}}\right), \quad W'_i=\frac 1{g_{,u}}W_i,
\quad \delta'_{ij}=\delta_{ij}.\nonumber
\eeq
\item{} $(v',u',x'^i)=(v,u,f^i(u;{x}^k))$, where $f^i(u_0;x^k)$ is an isometry of flat space, and $J^i_{~j}\equiv \frac{\partial f^i}{\partial x^j}$.
\beq
H'&=& H+\delta_{ij}f^i_{,u}f^j_{,u}-W_j\left(J^{-1}\right)^j_{~i}f^i_{,u}, \quad W'_i=W_j\left(J^{-1}\right)^j_{~i}-\delta_{ij}f^j_{,u}, \nonumber \\
 \delta'_{ij}&=&\delta_{kl}\left(J^{-1}\right)^k_{~i}\left(J^{-1}\right)^l_{~j}\equiv \delta_{ij}.\nonumber
\eeq
\end{enumerate}

Transformations (1) and (4) are both
subsets of the more general transformations (\ref{trans}) (diffeomorphisms
and isometries, respectively). The set of
transformations (4) therefore consists of translations and rotations
of the coordinates $x^i$.

The linearly independent components of the Riemann tensor with boost weight 1 and 0 are:
\beq
R_{ 0 1 0 (i+1)}&=&-\frac 12 W_{ i,vv}, \\
R_{ 0 1 0 1}&=& -H_{,vv}+\frac 14\left(W_{{i},v}\right)\left(W^{ i,v}\right), \\
R_{ 0 1 (i+1) (j+1)}&=& W_{[ i}W_{ j],vv}+W_{[ i; j],v}, \\
R_{ 0 (i+1) 1 (j+1)}&=& \frac 12\left[-W_{ j}W_{ i,vv}+W_{ i; j,v}-\frac 12
\left(W_{ i,v}\right)\left(W_{ j,v}\right)\right].
\eeq
Hence, the negative boost order conditions of the Riemann tensor yield
\beq
W_{ i,vv} &=& 0, \label{Wcsi1}\\
H_{,vv}-\frac 14\left(W_{ i,v}\right)\left(W^{ i,v}\right) &=& 0,
\eeq from which it follows that \beq W_{ i}(v,u,x^k)=v{W}_{
i}^{(1)}(u,x^k)+{W}_{ i}^{(0)}(u,x^k), \label{Weq}\eeq  \beq
H(v,u,x^k)=\frac{v^2}{8}({W}_i^{(1)})({W}^{(1)i})+v{H}^{(1)}(u,x^k)+{H}^{(0)}(u,x^k),\label{Heq}
\eeq subject to \beq W_{[ i; j],v} = 0,  \label{curl}\eeq and \beq
W_{( i; j),v}-\frac 12 \left(W_{ i,v}\right) \left(W_{ j,v}\right)
= 0.\label{sym}\eeq All of these spacetimes are VSI. Further
constraints can be made, and the resulting spacetimes invariantly
classified.

From eqn. (\ref{curl}) it follows that $W^{(1)}_i$ can be written
locally as a gradient: \beq W^{(1)}_i = [\phi(u, x^k)]_{,i}. \eeq
Eq. (\ref{sym}) now simplifies to
\beq
\left(e^{-\frac 12\phi}\right)_{,ij}=0,
\eeq
which can be integrated to yield:
\beq
\phi=-2\ln\left[a_i(u)x^i+C(u)\right],
\eeq
where $a_i(u)$ and $C(u)$ are arbitrary functions of $u$. 
Now, utilizing the rotations and translations of $x^i$ we can simplify $\phi$:
\beq
\phi=-2\ln[a(u)x^1], \quad \text{or} \quad \phi=-2\ln[C(u)].
\eeq
(Note that rotations in the subspace orthogonal to $x^1$ are still permitted).
Hence, we obtain the two general cases:
\beq
\text{(i):}&& W^{(1)}_1 = -\frac{2}{x^1};\quad W^{(1)}_i = 0,~ i\neq 1.\nonumber \\
\text{(ii):} && W^{(1)}_i = 0.\nonumber
\label{coords}
\eeq
The form of $H$ is given by eqn. (\ref{Heq}),
where ${H}^{(1)}(u,x^k)$ and ${H}^{(0)}(u,x^k)$ are the redefined
functions. The spacetimes above are in general of {\em Ricci and Weyl type III}.

Further progress can be made by classifing the metric in terms of their Weyl-type (III, N or O) and their Ricci type (N
or O), and the form of $L_{ab}$. In particular, ${H}^{(1)}(u,x^k)$ is determined in terms of the functions ${W}_{ i}^{(0)}(u,x^k)$ by restricting the metrics to be of Ricci type N. Additional constraints on $\Psi_{ij}$, $\Psi_{ijk}$, etc. can be obtained by employing
the Bianchi and Ricci identities \cite{bianchi}. In addition, the spatial tensors 
(indices $i,j$) can be written in
canonical form and remaining coordinate and frame freedom utilized.

\subsection{Kinematics}

The nonzero `spin coefficients' $L_{ij}$ are defined by \beq
\ell_{(a;b)} = L \ell_{a}\ell_{b} + L_i(\ell_{a} m^{i}_{b} +
\ell_{b} m^{i}_{a}). \eeq $L=L_{11}$ is the analogue of the spin
coefficient $(\gamma + {\bar{\gamma}})$ in 4-dimensions and the
$L_i \equiv L_{1i}$ are the analogues of the spin coefficient
$\tau$. From eqn. (\ref{Kundt}) we have that \beq L = H_{,v}, \eeq
and \beq L_i = \frac{1}{2} W_{i,v} = \frac{1}{2} W^{(1)}_{i}. \eeq
Therefore, in the preferred frame (\ref{coords}) $L_1 \neq 0$ and
all other $L_i$ are zero (i.e., the rotation used above also
simultaneously normalizes the $L_i$). All of the information from
the `spin coefficients' can then be summarized by: \beq W^{(1)}_1
= -2\frac{\epsilon}{x^1};\quad W^{(1)}_n = 0, n =2, ..., N-2, \eeq where
$\epsilon=0$ corresponds to $W^{(1)}_i = 0$ (and to $\tau=0$ in
4-dimensions) and $\epsilon=1$ in the case $L_1 \neq 0$ ($\tau
\neq 0$). Note that for $\epsilon=0$ and $H_{,v}=0$, $\ell_{a}$
is a null Killing vector.

\subsection{Ricci type}
The Ricci tensor is given by \beq R_{ab} =  \Phi \ell_{a}\ell_{b}
+ \Phi_{i} (\ell_{a} m^{i}_{b} + \ell_{b} m^{i}_{a}). \label{ricci} \eeq The Ricci
type is $N$ if $\Phi_{i}=0 = R_{1i}$ (otherwise the
Ricci type is III; Ricci type O is vacuum).  The non-zero components can be
simplified and chosen to be constant.

Further progress is most easily made by first performing a type
(2) coordinate transformation with $h(u,x^{k})$ satisfying
\begin{equation}
h_{,1}-2\frac{h\epsilon}{x^{1}}=W^{(0)}_{1}.
\end{equation}
The effect is to transform away $W^{(0)}_{1}$, so that in the new
coordinates $W_{1}=-2v\epsilon/x^{1}$ and the remaining metric
functions $W_{m}=W^{(0)}_{m}(u,x^{k})$, $H^{(1)}(u,x^{k})$ and
$H^{(0)}(u,x^{k})$ undergo a redefinition\footnote{\;It is also possible
to choose a gauge with $W^{(0)}_{1}$ non-zero; 
the corresponding results will be presented elsewhere \cite{proc}.}. That is, \beq
W_{1}=-2\frac{v\epsilon}{x^{1}}, \eeq and \beq
W_{m}=W^{(0)}_{m}(u,x^{k}), \eeq where $m$ ranges over
$2,\ldots, N-2$ (i.e., does not include $i=1$).

When $\epsilon=0,1$ the Ricci type N conditions $\Phi_{i}=0$ reduce to
\begin{eqnarray}
2H^{(1)}_{\quad,1} & = & \frac{2\epsilon}{x^{1}}W^{(0)m}_{\quad\quad,m}-W^{(0)m}_{\quad\quad,m1}  \label{rn1} \\
2H^{(1)}_{\quad,n} & = & \Delta W^{(0)}_{n}-W^{(0)m}_{\quad\quad,mn}  \label{rn2}
\end{eqnarray}
subject to
\begin{eqnarray}
\Delta W^{(0)}_{n,1} & = & \frac{2\epsilon}{x^1}W^{(0)m}_{\quad\quad,mn}   \label{icrn1} \\
\Delta W^{(0)}_{m,n} & = & \Delta W^{(0)}_{n,m}     \label{icrn2}
\end{eqnarray}
where $\Delta=\partial^{i}\partial_{i}$ is the spatial Laplacian and $m,n\geq 2$.  A partial integration of
(\ref{rn1})-(\ref{icrn2}) reduces these constraints to a divergence and a Laplacian that must be satisfied by
$W^{(0)}_{m}$, namely
\begin{eqnarray}
W^{(0)m}_{\quad\quad,m} & = & \epsilon(x^1)^{2}\left[F-\int\frac{4}{(x^1)^3}H^{(1)}dx^1 \right]
+ (1-\epsilon)F-2H^{(1)} \label{divrn} \\
\Delta W^{(0)}_{n} & = & \epsilon(x^1)^{2}\left[F_{,n}-\int\frac{4}{(x^1)^3}H^{(1)}_{\quad,n}dx^1 \right]
+(1-\epsilon)F_{,n} \label{lapvrn}
\end{eqnarray}
where $F=F(u,x^n)$ is an arbitrary function independent of $x^1$.  Note when $\epsilon=0$ (\ref{divrn}) defines
$H^{(1)}$ with $F$ determined from (\ref{lapvrn}). We note that, in general, $F$ cannot be
transformed away.

{\em Therefore, the equations above describe all $VSI$ metrics of
Ricci type N, with $\epsilon=0$ or $\epsilon=1$. In general the
Weyl tensor is of type III.}

\subsection{Weyl-type}

The Weyl tensor can be expressed as \beq C_{abcd} = 8 \Psi_{i}
\ell_{\{a} n_b \ell_{c} m^{i}_{d\}} + 8 \Psi_{ijk} m^{i}_{\{a}
m^{j}_b \ell_c m^{k}_{d\}} + 8 \Psi_{ij} \ell_{\{a} m^{i}_b
\ell_{c} m^{j}_{d\}} . \label{WeylIIIN} \eeq The case $\Psi_{ijk}
\not= 0$ is of Weyl type III, while $\Psi_{ijk} = 0$ (and
consequently also $\Psi_{i} = 0$) corresponds to type N. Note that
$\Psi_{ij}$ is symmetric and traceless. $\Psi_{ijk}$ is
antisymmetric in the first two indices with $\Psi_i=2 \Psi_{ijj}$, and in vacuum also
satisfies \beq  \Psi_{\{ijk\}}=0. \label{relijj}
\eeq

\subsubsection{Type III}

As noted above, in general the Weyl tensor is of type III. Let us treat first the case $\epsilon=1$:
\begin{eqnarray}
W_1 & = & -\frac{2}{x^1}v  \label{e1w1III} \\
W_{m} & = & W^{(0)}_{m}(u,x^{k})  \label{e1wmIII} \\
H & = & H^{(0)}(u,x^{i})+\frac{1}{2}\left(\tilde{F}-W^{(0)m}_{\quad\quad,m}\right)v+\frac{v^2}{2(x^1)^2}, \label{e1hIII}   
\end{eqnarray}
where $\tilde{F}=\tilde{F}(u,x^i)$ is a function satisfying:
\begin{equation}
\tilde{F},_1=\frac{2}{x^1}W^{(0)m}_{\quad\quad,m},\;\;\;\;\tilde{F},_n=\Delta W^{(0)}_{n}  \label{Feps1}
\end{equation}
In addition, we have the Einstein equation
\begin{eqnarray}
x^1 \triangle \left( \frac{H^{(0)}}{x^1} \right) &+& \left( \frac{W^{(0)m}W^{(0)}_m}{x^1}\right),_{1}
-2H^{(1)},_{m} W^{(0)m} -H^{(1)} W^{(0)m},_{m} 
\nonumber \\  & - &\frac{1}{4}W_{mn}W^{mn}- W^{(0)m}_{\quad\quad,mu}+\Phi=0,\label{Ruueps1}
\end{eqnarray}
where $W_{mn}=W^{(0)}_{m,n}-W^{(0)}_{n,m}$ 
\footnote{\;Note that, in general, $W^{(0)}_{1}=0$ but $W_{1n}=-W_{n1} \neq0$.} 
and $\Phi$ is determined by the matter field (see equation (\ref{ricci})); in the case of vacuum we have 
$\Phi=0$.

We consider now the case $\epsilon=0$:
\begin{eqnarray}
W_{1} & = & 0,  \label{e0w1III} \\
W_{m} & = & W^{(0)}_{m}(u,x^{k}) \label{e0wmIII} \\
H & = & H^{(0)}(u,x^{i})+\frac{1}{2}\left(F-W^{(0)m}_{\quad\quad,m}\right)v, \label{e0hIII}    
\end{eqnarray}
where $F=F(u,x^n)$ is a function satisfying:
\begin{equation}
F,_1=0,\;\;\;\;F,_n=\Delta W^{(0)}_{n} \label{Feps0}
\end{equation}

Finally we have:
\begin{eqnarray}
\triangle H^{(0)}&- &\frac{1}{4}W_{mn}W^{mn} 
-2H^{(1)},_{m} W^{(0)m} -H^{(1)} W^{(0)m},_{m} 
\nonumber \\  & - & W^{(0)m}_{\quad\quad,mu}+\Phi=0,
\label{Ruueps0} 
\end{eqnarray}
\\
$VSI$ spacetimes with 
(\ref{e1w1III})-(\ref{e1hIII}) and (\ref{e0w1III})-(\ref{e0hIII}) 
are the higher dimensional 
analogues of the four-dimensional spacetimes of Petrov (Weyl) type III,
PP-type O with $\tau \neq 0$ and $\tau=0$, respectively.

\subsubsection{Type III(a)}
Further subclasses can be considered \cite{class}. We discuss the subclass III(a) 
which is defined by vanishing $C_{011(n+1)}$. In the $\epsilon=1$ case
\begin{eqnarray} C_{0112}&=&H^{(1)},_{1} \nonumber \\ C_{011(m+1)}&=&H^{(1)},_{m}-\frac{1}{2}\frac{W^{(0)}_{m,1}}{x^1},  \label{iiiae1} \end{eqnarray}
where $H^{(1)}$ is the coefficient of $v$ in (\ref{e1hIII}) and $m=2,\ldots,N-2$. 
Vanishing of the Weyl components (\ref{iiiae1}) implies $H^{(1)},_{m1}=0$ for 
consistency; these constraints give already
\beq
W^{(0)}_{m}  =  \tilde{W}_m(u,x^n)\;\frac{(x^1)^2}{2}+\tilde{\tilde{W}}_m(u,x^n), \label{wmIIIa}
\eeq 
where the functions $\tilde{W}_m$, $\tilde{\tilde{W}}_m$ do not depend on $x^1$. The Weyl components (\ref{iiiae1}) vanish themselves when the functions $W^{(0)}_m$ satisfy:
\begin{eqnarray} W^{(0)m}_{\quad\quad,m1}&=&\frac{2}{x^1}W^{(0)m}_{\quad\quad,m}
\label{cond1e1} \\ W^{(0)m}_{\quad\quad,mn} &=& \Delta W^{(0)}_{n}-\frac{W^{(0)}_{n,1}}{x^1}
\label{cond2e1} \end{eqnarray}
Inserting here (\ref{wmIIIa}) we get the following conditions for the functions $\tilde{W}_n(u,x^n)$, $\tilde{\tilde{W}}_n(u,x^n)$: 
\begin{eqnarray}
\tilde{W}^{m}_{\quad,mn} &=& \Delta \tilde{W}_{n} \label{cond1e1} \\
\tilde{\tilde{W}}^{m}_{\quad,m}&=&\Delta \tilde{\tilde{W}}_{n}=0 \label{cond2e1}
\end{eqnarray}
The class is characterized by
\begin{eqnarray}
W_{1} & = & -\frac{2}{x^1}v,  \label{e1w1IIIa} \\
W_{m} & = & \tilde{W}_m(u,x^n)\;\frac{(x^1)^2}{2}+\tilde{\tilde{W}}_m(u,x^n), \label{e1wmIIIa} \\
H & = & H^{(0)}(u,x^{i})+\frac{1}{2}f(u,x^n)\;v+\frac{v^2}{2(x^1)^2} , \label{e1hIIIa}    
\end{eqnarray} 
where the function $f(u,x^n)$ does not depend on $x^1$ and it is such that $f,_{m}=\tilde{W}_m(u,x^n)$. A necessary (but not sufficient) condition 
for a metric to be in this subclass is that the functions 
$\tilde{W}_{m}$, $\tilde{\tilde{W}}_m$ satisfy equations (\ref{cond1e1}), 
(\ref{cond2e1}). Furthermore, they have to be such that (some of) 
the boost weight $-1$ ~~$C_{1ijk}$ Weyl components are 
non-vanishing{\footnote {\;There is no compact way to write down the $C_{1ijk}$ in a 
similar fashion to that in the III(a) 
$\epsilon=0$ case (equation (\ref{cond3})), so we shall not explicitly display them here.}} 
in order for the considered metric to be truly of type III(a).

An example of a spacetime in this subclass is constructed as follows. For odd 
$N$, consider the functions
\begin{eqnarray}
\tilde{\tilde{W}}_n=\frac{-p_n(u)x^{n+1}}{Q^{\frac{N-3}{2}}}, \\
\tilde{\tilde{W}}_{n+1}=\frac{p_n(u)x^{n}}{Q^{\frac{N-3}{2}}},
\end{eqnarray}    
where $n$ only takes on even values of $m$ and $Q=\delta_{mm'}x^mx^{m'}$. 
The $p_n$ are arbitrary functions of $u$. For even $N$, consider the same functions except for $n=N-2$:
\beq
\tilde{\tilde{W}}_{N-2}=0  
\eeq
Note that none of these functions depend on $x^1$, as required. 
The spacetimes are then characterized by equations (\ref{e1w1IIIa})-(\ref{e1hIIIa}), 
with $\tilde{W}_m(u,x^n)=f(u,x^n)=0$. The choice of functions $W_m$ in $N$ odd (even)
dimensions coincides with those of \cite{frolov} for $N-1$ even (odd) dimensions. 
However, these spacetimes are more general than the relativistic gyratons, 
since the functions $W_1$ and $H$ can depend on $v$; 
these solutions might be referred to as {\em Kundt gyratons}, in analogy with the pp-waves
(no $v$-dependence) and 
Kundt waves ($v$-dependence) but which have the same (vanishing) $W_m$ functions. 

Another example of a spacetime in this subclass with $\tilde{W}_m=0$ is given by the six-dimensional metric:
\begin{eqnarray} W_{1} & = & -\frac{2}{x^1}v,  \\ W_{2} & = &
f_2(u)x^3x^4,  \\ W_{3} & = &
f_3(u)x^2x^4, \\ W_{4} & = &
f_4(u)x^2x^3, \\ H & = & H^{(0)}(u,x^{i})+\frac{v^2}{2(x^1)^2}, 
\end{eqnarray}

In the final type III(a) example, $H$ contains a linear dependence
on $v$ (the two examples above can be generalized to include a linear 
dependence on
$v$ as well).  From equations (\ref{e1w1IIIa}) - (\ref{e1hIIIa}), we consider the
symmetric and antisymmetric parts of $\tilde{\tilde{W}}_{m,n}$.  We impose the condition
\begin{equation}
\left(\tilde{\tilde{W}}_{m,n}-\tilde{\tilde{W}}_{n,m}\right)_{,l}=0,  \label{asymh}
\end{equation}
\noindent along with the constraint (\ref{cond2e1}) and recall that $f,\tilde{W}_{m}$ and
$\tilde{\tilde{W}}_m$ contain no $x^{1}$ dependence. This results in the vanishing
of all components,
$C_{011(n+1)}$, of the Weyl tensor, whereas the remaining boost weight -1 components, 
$C_{1ijk}$, are related to
the symmetric part of $\tilde{\tilde{W}}_{m,n}$.  We note that, if instead of 
(\ref{asymh}), we required that
$\tilde{\tilde{W}}_{(m,n)}=0$ (and also (\ref{cond2e1})), the resulting 
Weyl tensor is of reduced type N.

In the $\epsilon=0$ case
\begin{equation} C_{011(n+1)}=H^{(1)},_{n} \label{iiiae0} \end{equation} where $H^{(1)}$ is the coefficient of $v$ in (\ref{e0hIII}) and $n=1,\ldots,N-2$. The Weyl components (\ref{iiiae0}) vanish when the functions $W^{(0)}_m$ satisfy:  \begin{eqnarray} W^{(0)m}_{\quad\quad,m1}&=&0
\label{cond1} \\ W^{(0)m}_{\quad\quad,mn} &=& \Delta W^{(0)}_{n}
\label{cond2} \end{eqnarray} 
These conditions can be written more compactly:
\beq
\partial_m W^{mn}=0 \label{cond1,2}
\eeq
(and are consequently expressed in a form similar to the Maxwell equations in Euclidean space). 
The following class is obtained:
\begin{eqnarray} W_{1} & = & 0, \label{brinkw1} \\ W_{m} & = &
W^{(0)}_{m}(u,x^{k}) \label{brinkwm} \\ H & = & H^{(0)}(u,x^{i}), \label{brinkh0}
\end{eqnarray} 
A necessary (but not sufficient) condition for a metric to be in this subclass is that the functions $W_{m}$ satisfy equations (\ref{cond1}), (\ref{cond2}). Furthermore, the functions $W_{m}$ have to be such that (some of the) boost weight $-1$ Weyl components are non-vanishing:
\be
C_{1(l+1)(n+1)(m+1)}=\frac{1}{2} \left( W_{m,n}-W_{n,m} \right),_l \neq 0\label{cond3}
\ee
This is equivalent to:\beq  \partial_l W^{mn} \neq 0. \eeq
In general, there is no Lorentz transformation such that these  boost weight $-1$  
components can be set to zero; hence (\ref{brinkw1})-(\ref{brinkh0}) 
is in general of Weyl type III (and not of type N; i.e., in this subclass
$\Psi_{i}=0$ does not imply $\Psi_{ijk} = 0$).
The higher dimensional relativistic gyraton spacetime in \cite{frolov} is an 
example of a spacetime in this subclass. A five-dimensional example of another spacetime in this subclass is given by:
\begin{eqnarray} W_{1} & = & 0,  \\ W_{2} & = &
f_2(u)x^1x^3,  \\ W_{3} & = &
f_3(u)x^1x^2, \\ H & = & H^{(0)}(u,x^{i}), 
\end{eqnarray}  
Here, $f_2$ and $f_3$ are arbitrary functions of $u$. We note that metrics in this subclass have a covariantly constant 
null vector (see later).

\subsubsection{Type N}

The spacetime is of Weyl type N if: \beq \Psi_{ijk} = 0 =
C_{1kij}, \Psi_{i} = 0 = C_{101i}. \label{wtypen} \eeq Further
constraints on $\Psi_{ij}$, $\Psi_{ijk}$, etc. can be obtained by
employing the Bianchi and Ricci identities.

Further progress can then be made by requiring the vanishing of
boost weight -1 components of the Weyl tensor; using the above
results and equations (\ref{divrn}) and (\ref{lapvrn}), we obtain
the following metric functions. \\ 
{\em{Generalized Kundt waves}}. For $\epsilon=1$ we have:
\begin{eqnarray}
W_{1} & = & -2\frac{v}{x^{1}}, \label{eps1w1} \\
W_{m} & = & x^{n}B_{nm}(u)+C_{m}(u),  \label{eps1wm}\\
H & = & \frac{v^2}{2(x^{1})^{2}}+H^{(0)}(u,x^{i}).   \label{eps1h}
\end{eqnarray}
And (\ref{Ruueps1}) simplifies to
\begin{equation}
x^1 \bigtriangleup \left( \frac{H^{(0)}}{x^1} \right)-\frac{1}{(x^1)^2}\sum W_m^2 -2 \sum_{m<n} B_{mn}^2+\Phi=0, \label{Ruueps1N}
\end{equation}
where $W_m$ is given by (\ref{eps1wm}). In the case $B_{nm}=C_m \equiv 0$ one obtains the higher dimensional Kundt waves in \cite{CMPPPZ}. There are further subclasses in which further simplification
occurs (in which certain components can be set to zero, and others
can be set constant).

\noindent {\em Generalized pp-waves}.
In the case $\epsilon=0$ we have
\begin{eqnarray}
W_1 & = & 0 \label{eps0w1} \\
W_{m} & = & x^1 C_m(u)+ x^{n}B_{nm}(u),   \label{eps0wm} \\
H & = & H^{(0)}(u,x^{i}),    \label{eps0h}
\end{eqnarray}
and (\ref{Ruueps0}) reduces to:
\begin{equation}
\triangle H^{(0)}-\frac{1}{2}\sum C_m^2 -2 \sum_{m<n} B_{mn}^2+\Phi=0, \label{Ruueps0N}
\end{equation}
$B_{nm}=B_{[nm]}$ in (\ref{eps1wm}) and (\ref{eps0wm}). In (\ref{eps0wm}) a type (2) coordinate transformation has
been used to remove a gradient term and in (\ref{eps0h}) a type (3) transformation was used to eliminate the linear $v$
dependence. As in the four-dimensional case the term $x^1 C_m(u)$ can be transformed
away (at the expense of introducing a non-vanishing $W_1$).
However, unlike the four-dimensional case, terms
linear in $x^{n}$ in $W_{m}$ cannot be transformed away (even if a 
non-zero $W_1$ is allowed).



Since both Weyl and Ricci tensors have only boost weight -2 components (i.e. both 
are of type N), $VSI$ spacetimes with 
(\ref{eps1w1})-(\ref{eps1h}) and (\ref{eps0w1})-(\ref{eps0h}) 
are the higher dimensional generalizations of Kundt and pp-waves (i.e., are the 
higher dimensional
analogues of the four-dimensional spacetimes of Petrov (Weyl) type N,
PP-type O with $\tau \neq 0$ and $\tau=0$, respectively). 
Higher dimensional  pp-wave spacetimes have been studied extensively
\cite{matsaev,hortseyt,tseytlin,RT,russot}.
 
\subsubsection{Type O}

The spacetime is of type O if the Weyl tensor vanishes. 
We consider the case where $\Phi=\Phi(u)$ in equation (\ref{ricci}). \\

For $\epsilon=1$,  the function $H^{(0)}$ must satisfy:  \beq
x^1\left(\frac{H^{(0)}}{x^1}\right),_{11}=\frac{1}{(x^1)^2}\sum W_m^2
-\frac{1}{8}\Phi,\;\;\;H^{(0)},_{mm}=\sum_{n} B_{mn}^2-\frac{1}{8}\Phi \eeq It can be seen
that this cannot be accomplished in the case $\Phi=\Phi(u)$. However, the Weyl tensor does 
vanish for $\Phi=\Phi_0(u)x^1$ and  
\beq \;\;\;\;\;\;H^{(0)}=\frac{1}{2}\sum W_m^2
-\frac{1}{16}\Phi_0(u)x^1 \left[(x^1)^2+(x^m)^2\right] +x^1 F_0(u)+x^1x^iF_{i}(u).
\label{weylOeps1} \eeq Here $F_0(u)$, $F_i(u)$ are arbitrary functions of $u$.  The last two
terms in $H^{(0)}$ correspond to the Weyl type O reduction of the higher dimensional Kundt waves. This particular form of $\Phi$ is likely to be the most general one compatible
with a vanishing Weyl tensor.  It remains to clarify the associated type of matter field
for this conformally flat spacetime and the given Ricci tensor.
Null Maxwell fields and massless scalar fields  can already be excluded, as is the case
in four dimensions \cite{edgar-ludw}.  \\

The case $\epsilon=0$:  \begin{eqnarray} H^{(0)}&=&
\frac{1}{8}\left( \sum C^2_m (x^1)^2 + \sum_{m \leq n}C_m C_n x^m
x^n\right)+\frac{1}{2}x^1x^m(C'_m+B_{mn}C_n) \nonumber \\ & & + \frac{1}{2}B_{ml}B_{nl}x^m x^n
 +x^i F_i(u)-\frac{1}{16}\Phi\;\left[(x^1)^2+(x^m)^2\right].  \label{HtypeOeps0} \end{eqnarray} 
Here $'$ means derivative with respect to $u$ and $l \neq
m,n$; $F_i(u)$ are arbitrary functions of $u$.  The term $x^i F_i(u)$ corresponds to the
Weyl type O reduction of generalized pp-waves with $W_m=0$. In the case of matter depending on
the spatial coordinates as well, $\Phi(u,x^i)$, the last term on the 
right-hand side  of  
equation (\ref{HtypeOeps0}) has to be replaced by some $H^0_{\Phi}$ such that:  \beq
H^0_{\Phi,_{ii}}=-\frac{1}{8}\Phi,\;\;\;H^0_{\Phi,_{ij}}=0,\;\;\;\;\;i,j=1,\ldots,8 . \label{HtypeOeps02} \eeq 

\subsection{Ricci type O - vacuum} 
In the vacuum case and Weyl type III or N, equations are the same as for 
Ricci type N except $\Phi$ is now zero. Finally, in the case of Weyl type O, we simply have N-dimensional Minkowski space.


\section{Discussion}

In Table $1$ we collect together the results of our analysis, 
listing the metric functions (and any remaining constraints for the 
higher dimensional $VSI$ spacetimes) according to the algebraic classification
of the Weyl tensor.

\begin{table}[h] \label{table:1}
\begin{center}
\scriptsize{
\begin{tabular}{|l|c|c|} 
\hline
&&\\
$\epsilon$ & {\bf Weyl type} & {\bf Metric functions}  \\
&&\\
\hline
&&\\

& III & $\begin{array}{c} W_{1}  = 0 \\  
W_{m}  =  W^{(0)}_{m}(u,x^{k})  \\
H  =  H^{(0)}(u,x^{i})+\frac{1}{2}\left(F-W^{(0)m}_{\quad\quad,m}\right)v;\;\;
F(u,x^i)\;\mbox{defined by}\;(\ref{Feps0}) \\ \\ \mbox{Eq.}\; (\ref{Ruueps0}) \end{array}$ \\
&&\\
\cline{2-3}
&&\\
0 & N &  $ \begin{array}{c} \mbox{{\em Generalized pp waves:}} \\ \\ W_1  =  0  \\
W_{m} =  x^1 C_m(u)+ x^{n}B_{nm}(u)   \\
H  =  H^{(0)}(u,x^{i}) \\ \\ \mbox{Eq.}\;(\ref{Ruueps0N}) \end{array}$ \\
&&\\
\cline{2-3}
&&\\
& O & $W_1$, $W_m$ as in type N; $H^{(0)}$ given by (\ref{HtypeOeps0}), (\ref{HtypeOeps02}) \\
&&\\
\hline
&&\\ 
& III & $\begin{array}{c} W_1  =  -\frac{2}{x^1}v  \\
W_{m}  =  W^{(0)}_{m}(u,x^{k}) \\
H  =  H^{(0)}(u,x^{i})+\frac{1}{2}\left(\tilde{F}-W^{(0)m}_{\quad\quad,m}\right)v+\frac{v^2}{2(x^1)^2};
\;\; \tilde{F}(u,x^i)\;\mbox{defined by}\;(\ref{Feps1}) \\ \\ \mbox{Eq.}\; (\ref{Ruueps1}) \end{array}$ \\
&&\\
\cline{2-3}
&&\\
1 & N &  $\begin{array}{c} \mbox{{\em Generalized Kundt waves:}} \\ \\ W_{1}  =  -2\frac{v}{x^{1}} \\
W_{m}  =  x^{n}B_{nm}(u)+C_{m}(u) \\
H  =  \frac{v^2}{2(x^{1})^{2}}+H^{(0)}(u,x^{i}) \\ \\ \mbox{Eq.}\;(\ref{Ruueps1N}) \end{array}$ \\
&&\\
\cline{2-3}
&&\\
& O & $W_1$, $W_m$ as in type N;\; $H^{(0)}$ given by (\ref{weylOeps1}) and $\Phi=\Phi_0(u)x^1$ \\
&&\\
\hline
\end{tabular}
}
\caption{All higher dimensional $VSI$ spacetimes of Ricci type N. 
In the above, $m=2,\ldots,N-2$. Ricci type O (vacuum) spacetimes occur 
for $\Phi=0$ in (\ref{Ruueps1}), (\ref{Ruueps0}), (\ref{Ruueps1N}), (\ref{Ruueps0N}).}
\end{center}
\end{table}

\begin{table}[h] \label{table:2}
\begin{center}
\scriptsize{
\begin{tabular}{|l|c|c|c|} 
\hline
&&&\\
$\epsilon$ & {\bf Ricci type} & {\bf Weyl type} & {\bf Metric functions}  \\
&&&\\
\hline
&&&\\

& III & III & $\begin{array}{c} W_{1}  = 0 \\  
W_{m}  =  W^{(0)}_{m}(u,x^{k})  \\
H  =  H^{(0)}(u,x^{i}) \\ \\ 
-\triangle H^{(0)}+\frac{1}{4}W_{mn}W^{mn} + W^{(0)m}_{\quad\quad,mu}=
\Phi + \delta^{mn} W^0_m \Phi_{(n+1)} \;
\\ W^{(0)m}_{\quad\quad,m1}=2\Phi_2 \\ 
-\Delta W^{(0)}_{n}+W^{(0)m}_{\quad\quad,mn}=2\Phi_{n+1} \end{array}$ \\
&&&\\
\cline{2-4}
0 &&&\\
 & N & III(a) & $ \begin{array}{c}  W_1  =  0  \\
W_{m} =  W^{(0)}_{m}(u,x^{k})   \\
H  =  H^{(0)}(u,x^{i}) \\ \\ \mbox{Eq.}\; (\ref{Ruueps0}), \; \left(H^{(1)}=0\right) \\  
\\ W_m \;\mbox{satisfy}\; (\ref{cond1}),\;(\ref{cond2}),\;
(\ref{cond3}) \end{array}$ \\
&&& \\
\cline{3-4}
&&&\\
&  & N & {\em Generalized pp waves}; \; Eqns. (\ref{eps0w1}) - (\ref{Ruueps0N})  \\
&&&\\
\hline

\end{tabular}
}
\caption{All higher-dimensional $VSI$ spacetimes with a covariantly 
constant null vector are presented. The $\Phi_i$, where $i=2,...N-1$, are
given in equation (\ref{ricci}).}
\end{center}
\end{table}

In higher dimensions the spacetimes with vanishing zeroth order curvature invariants,
called VSI$_{0}$, have Ricci and Weyl type III, N or O \cite{Higher}.  Spacetimes
with vanishing zeroth and first order invariants, called VSI$_{1}$ spacetimes, 
were discussed in the case
of four dimensions in \cite{epsilon}.  It is plausible that in
higher dimensions the proper VSI$_{1}$ spacetimes (i.e.,  not VSI), have
Weyl type N, Ricci type N or O, and admit an aligned geodesic null congruence.
More specifically, consider the null frame $e_{a}=\{ \ell,n,m_{i}\}$ where
$i=2,\ldots,N-2$. Then assuming the aligned geodesic null congruence has been 
affinely parametrized, we can write its covariant derivative \cite{Higher} as
$\ell_{\alpha;\beta}=L_{11}\ell_{\alpha}\ell_{\beta}+L_{1i}\ell_{\alpha}m^{i}_{ \
\beta}+L_{i1}m^{i}_{\ \alpha}\ell_{\beta}+L_{ij}m^{i}_{\ \alpha}m^{j}_{\ \beta}$.
A decomposition of $L_{ij}$ into symmetric and antisymmetric parts gives the
expansion, shear, and twist of $\ell$.  The VSI$_{ 1}$ subclasses will arise from
the vanishing of certain combinations of these optical scalars. However, unlike the four dimensional result \cite{epsilon}, VSI$_{1}$
spacetimes
do not exist in the higher dimensional Robinson-Trautman class
\cite{bianchi,Jiri}. 
We expect the
corresponding metrics to be contained in the higher dimensional analogue of the
Pleba\'{n}ski class along with a class
containing the higher dimensional Hauser solution \cite{kramer}.

As noted earlier, the aligned, repeated, null vector $\ell$ of (\ref{Kundt}) is a null Killing
 vector (KV) if and only if $H_{,v}=0$ and
$W_{i,v}=0$ ($\epsilon=0$) (whence the
 metric no longer has any $v$
dependence).  Furthermore, since $L_{AB}:=\ell_{A;B}=\ell_{(A;B)}$ it follows 
that in this case if $\ell$ is a null KV then it
is also covariantly constant.  Without any further restrictions, the higher 
dimensional $VSI$ metrics
(\ref{Kundt}) admitting a null KV have Ricci and Weyl type III. The $VSI$ spacetimes
are recurrent if $\epsilon=0$ in $\ell$.

Therefore, in higher dimensions a $VSI$ spacetime which admits a covariantly constant
null vector (CCNV) has a metric of the form (\ref{brinkw1})-(\ref{brinkh0}) 
and, in general, is of Ricci type III and Weyl type III. 
These $VSI$ spacetimes are summarized in Table $2$ 
(however, these spacetimes are not the most general spacetimes with a CCNV; 
see \cite{CCNV}). The subclass of
Ricci type N  CCNV spacetimes are 
related to the ($F=1$) chiral null models (satisfying
eqn. (2.14) subject to eqns. (2.15)-(2.16))  of \cite{hortseyt}. 
The subclass of
Ricci type N and Weyl type III(a) spacetimes in which the
functions $W_{m}$ in (\ref{brinkwm}) satisfy the conditions (\ref{cond1}), 
(\ref{cond2}), (\ref{cond3}) includes the
relativistic gyratons \cite{frolov}, as was noted above.
The subclass of Ricci type N and Weyl type N spacetimes are
the {\em generalized pp-wave spacetimes}. These are the analogues of the four-dimensional
Brinkmann pp-waves \cite{brink} (which are always of PP-type O (Ricci type
N) and Petrov (Weyl) type N).
We emphasise that, unlike in four-dimensions, {\em in higher dimensions
$VSI$ spacetimes with a CCNV are, in general, of Ricci type III and Weyl type III}.

It is the  higher dimensional (generalized) pp-wave spacetimes that have been most
studied in the literature.
It is known that such spacetimes are exact solutions in string
theory \cite{string}. 
Recently, type-IIB superstrings in pp-wave
backgrounds with an ~R-R five-form field were also shown to be
exactly solvable \cite{matsaev}. In the context of string theory,
higher dimensional generalizations of pp-wave  backgrounds have
been considered \cite{hortseyt,tseytlin}, including string models
corresponding not only to the NS-NS but also to certain R-R
backgrounds \cite{RT,russot}, and pp-waves in eleven- and
ten-dimensional supergravity theory \cite{GuevenAD}. In four-dimensions,
a wide range of $VSI$~spacetimes have been shown to be exact solutions in string
theory \cite{coley}.  The same is expected in higher dimensions. 
Indeed, $VSI$ supergravity solutions can be constructed
\cite{hortseyt,Frolov2}
(although attention has been mainly focussed on the plane wave versions
of generalized pp-waves), and it is 
likely that all $VSI$ spacetimes are solutions of 
superstring theory when supported by appropriate bosonic fields.
In addition, the $VSI$ CCNV metrics preserve supersymmetry 
when embedded in $\mbox{N}=1$, $\mbox{D}=10$ supergravity \cite{ssw}. 
Whether this is also the case for other supergravity theories and/or 
more general $VSI$ spacetimes  
remains to be studied. We shall return to this in future work.

----------------------

{\em Acknowledgements: This work was supported by NSERC (AC and NP),
the Killam Foundation and AARMS (SH) and the programme FP52 of the Foundation for Research of Matter, FOM (AF).

\end{document}